The purple *Codex Rossanensis*: spectroscopic characterization and first evidence of the use of the elderberry lake in a 6$^{th}$ century manuscript
Marina Bicchieri

Marina Bicchieri (corresponding author)
Laboratory of Chemistry
Istituto centrale restauro e conservazione patrimonio archivistico e librario (Icrcpal)
Via Milano 76, 00184 Roma (Italy)
e-mail: marina.bicchieri@beniculturali.it

**ABSTRACT**
This paper presents the results obtained during the measurements campaign started in June 2012 and ended in November 2013 on the invaluable Purple *Codex Rossanensis*, 6$^{th}$ century, one of the oldest surviving illuminated manuscripts of the New Testament. The spectroscopic analyses performed by micro-Raman, micro-Fourier Transform Infrared, and X-Ray Fluorescence allowed for the complete characterization of the pictorial palette, the inks, the support and the materials used in a previous restoration treatment executed in 1917-19 by Nestore Leoni, a famous miniaturist, active from the end of 19$^{th}$ century to mid 20$^{th}$ century. To the author's knowledge the article shows the first experimental evidence of the use of the elderberry lake in a 6$^{th}$ century illuminated manuscript.

**Introduction**
The *Codex Rossanensis* is a 6$^{th}$ century Byzantine illuminated manuscript written on purple parchment, conserved at the Museo Diocesano in Rossano Calabro (Cosenza, Italy).
It was found in 1879 in the sacristy of the Cathedral of Maria Santissima by Adolf von Harnack and published soon after by Oscar von Gebhardt [1]. It is a Gospel in Greek consisting of 188 parchment sheets (31 cm x 26 cm, with a posterior numbering of each page from 1 to 376). It contains the Gospel of Matthew and the Gospel according to Mark (the latter with one lacuna, Mark 16:14-20), as well as part of the letter from Eusebius to Carpianum on the concordance of the Gospels. Originally it seems that it should have contained all four canonical Gospels, as shown in the miniature with the symbols of the four Evangelists, and in particular by the presence of the Eusebian concordances. The written part is laid in biblical uncial capital in two columns.
The manuscript is famous for its prefatory cycle of 13 miniatures of subjects from the Life of Christ, arranged in two tiers on the page, the miniature of the four Evangelists, the golden decoration of the letter to Carpianum, the magnificent illumination of Mark inspired by the Sophia and for the use of the very precious purple dye as background for all the parchments, and gold and silver for the text.
In 1917-19 the Codex was subjected to a restoration treatment, carried out by Nestore Leoni, a famous miniaturist, active from the end of 19$^{th}$ century to mid 20$^{th}$ century. Leoni's intervention irreversibly modified the aspect of the illuminated sheets. Nestore Leoni never wrote which materials he used for the restoration.
In June 2012 the Codex arrived at the Istituto Centrale Restauro e Conservazione Patrimonio Archivistico e Librario (Icrcpal) of Rome, for a complete characterization of the pigments, the support and the materials used by Nestore Leoni, the state of conservation and for the restoration.
The laboratory of chemistry of Icrcpal performed spectroscopic analyses, by micro-Raman (378 spectra), micro-Fourier Transform Infrared (80 spectra) and X-Ray Fluorescence (35 spectra), on the whole volume, both on the pigments and on the support.
The challenge of the analysis of the *Codex Rossanensis* lies in the lack of analytical information on the pictorial media used in Early Middle Ages (4$^{th}$ - 9$^{th}$ centuries). Even though old-medieval illuminated manuscripts have been deeply studied from the historical standpoint, they have been rarely described in their material composition [2]. Moreover, a recently discovered medieval Arab manuscript containing recipes on inks and pigments manufacture [3] and the careful translation work carried out during a PhD Thesis [Sara Fani, *Studi sul Vicino Oriente e Maghreb. Specificità culturali e relazioni interculturali*, Università degli Studi di Napoli "L'Orientale"] on another Arab manuscript [4], never translated before, offered precious tools for the interpretation of the experimental data.
In this paper I present the results obtained during the measurements campaign on the *Codex* executed between June 2012 and November 2013.
For a better understanding of some spectra acquired on red lakes, laboratory samples were prepared, using historical lake samples either belonging to the collection of the Icrcpal chemistry laboratory, or newly synthesized.

**Materials and Methods**
*Instrumentation*
Measurements were performed by means of a Renishaw In-Via Reflex Raman microscope equipped with a Renishaw diode laser at 785 nm (nominal output power 300 mW) and a1200 line/mm grating to disperse the backscattered light.

The Raman signal is detected by a Peltier cooled (-70° C) deep depletion charge-coupled device (CCD RD-VIU, 578 x 384 pixel) optimized for near-infrared and ultraviolet. The nominal spectral resolution obtained for the measurements is about 3 cm$^{-1}$. The system, equipped with a Leica DM LM microscope to focus the laser on the sample and a colour video camera, allows for the positioning of the sample and the selection of a specific region for the investigation. Spectral acquisitions (1-10 accumulations, 50 s each) were performed with a 50x objective (N.A. 0.75). Under these conditions, the laser spot measures about 20 μm$^2$.
Depending on the sample investigated, the laser power has been reduced with neutral density filters up to 0.03 mW.
Micro-FTIR measurements were performed using a Nexus Nicolet interferometer and a Continuμm Microscope, equipped with a KBr beam splitter, a liquid nitrogen cooled MCT/A detector and an Infinity Reflachromat$^{TM}$ 15X ∞/V objective with N.A. = 0.58. Measurements on a surface of 100 x 100 μm$^2$ were performed in the 4000-650 cm$^{-1}$ range at a resolution of 8 cm$^{-1}$, averaging 200-400 acquisitions per sample. No ATR spectra were collected in order to avoid any direct contact with the manuscript during the analyses.
XRF spectra were recorded by means of an Assing Lithos 3000 portable spectrometer, equipped with a Mo X-ray tube. With such an instrument, the radiation can be collimated at different beam diameters (from 0.5 to 4 mm), depending on the area of interest. In this experiment, the 2 mm collimator was used together with a Zr filter. A red LASER (695 nm) and a camera (both integrated into the system and controlled by the instrument software) were used to choose the area to be sampled. Measurements were performed with the tube operating at 25 kV, 0.300 mA, in the 0-25 keV range with a resolution of 160 eV at 5.9 keV, lasting 10-60 minutes for each acquisition. Long-time acquisitions were performed in order to obtain information on the elements present in minor amount.

*Laboratory samples*
- Historical samples of lakes, belonging to the collection of the Icrcpal chemistry laboratory (gift of Lorilleux, Milano, 1938) obtained from *Rubia Tinctorum* (red) and *Porphyrophora hamelii* (Armenian cochineal, red) were prepared following the indication of Bisulca et al. [5], by using eggs: yolks and egg whites were separated and vigorously whisked. They were then combined by mixing 2:1:1 yolk: egg white: water and added to the pulverised lake, until the correct viscosity of the lake was obtained. The pigment was then applied on the surface of a parchment.
- Aluminium lake pigments from *Crocus Sativus* (red) and *Sambucus Nigra* (pink-violet) were prepared following the ancient recipes reported in [6] and in [7]. Each pigment was then mixed with egg white as reported in [8] recipe XVI and applied on the parchment.
- Two different kind of purple-dyed parchment samples were prepared, by treating the parchment with: a) aqueous solution of *Roccella Tinctoria* prepared as described in recipe 131 of the Stockholm papyrus [9], b) aqueous solution of *Roccella Tinctoria* and sodium carbonate, as described in the recipe 123 in [9].

**Results and discussion**
*Analysis of the materials used in the previous restoration (1917-19)*
On the first 20 pages of the codex, a layer of an unknown material was applied during the restoration performed in 1917-19 by Nestore Leoni, who never wrote any detail on the compounds used in the restoration, not even in the technical reports he had to present to the Ministry, to get paid for his work.
The unpublished reports are conserved at the Archivio Centrale dello Stato, Roma (Italy).
The materials used to reinforce the parchment sheets deeply penetrated into the bulk of the membranaceous support, modifying its optical characteristics: the sheets restored in this way are now completely transparent and their colour appears more brown than purple (Fig. 1-left). On some pages the applied layer was partially or completely detached (Fig. 1-right), as can be seen on page 8, where it was possible to analyse a fragment of the layer.

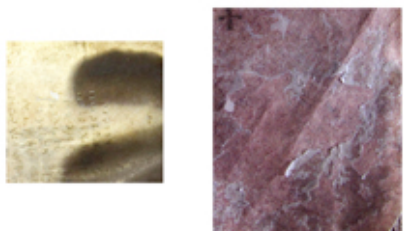

**Fig. 1 Appearance of the sheets treated by Nestore Leoni. Left: the parchment became transparent after the application of the reinforcing layer. Right: partially detached fragments of the applied layer are visible.**

By observing the detached layer under the Raman microscope, there was an evidence of the presence of some fibres, from which the spectra of cellulose had then been collected (Fig. 2). In the remaining part of the pellicle both Raman (Fig. 3) ad infrared analyses gave the unmistakeable spectrum of collagen, but the analyses carried out directly on other pages showed also the presence of cellulose nitrate, in some restricted regions (Fig. 4).
Three products were basically used between the end of 19$^{th}$ century and the beginning of 20$^{th}$ century for the reinforcement of damaged parchments: pure high quality gelatine mixed with formaldehyde, directly applied on the parchment or reinforced with Japanese paper, Archiv-Zapon (cellulose nitrate dissolved in amyl acetate with addition of camphor or formaldehyde), and Cellit (cellulose acetate mixed with acetic ether, ethanol, acetic acid and camphor) [10-16]. No traces of cellulose acetate were found in the codex.

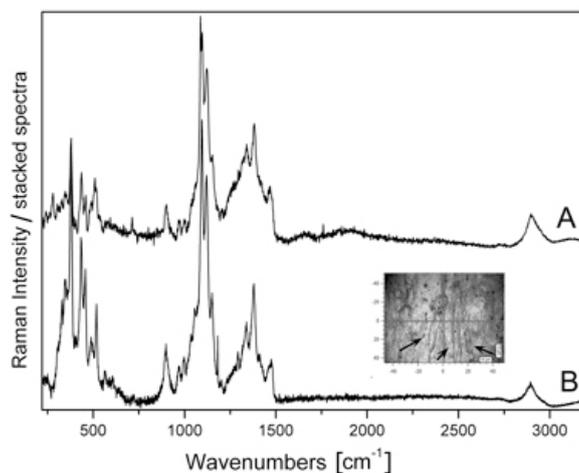

**Fig. 2 Raman spectra collected from A: fibres included in the restoration layer (marked with arrows in the inset) and, for comparison, B: spectrum of microcrystalline cellulose**

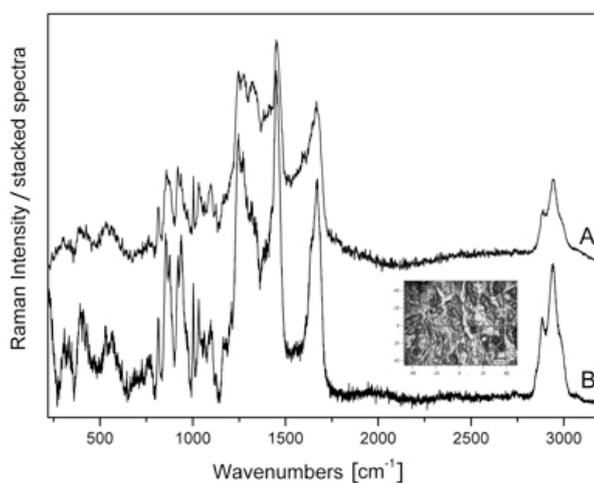

**Fig. 3 Raman spectra of A: restoration layer made of gelatine, compared to B: standard sample of collagen. The inset shows the appearance of the layer under the Raman microscope**

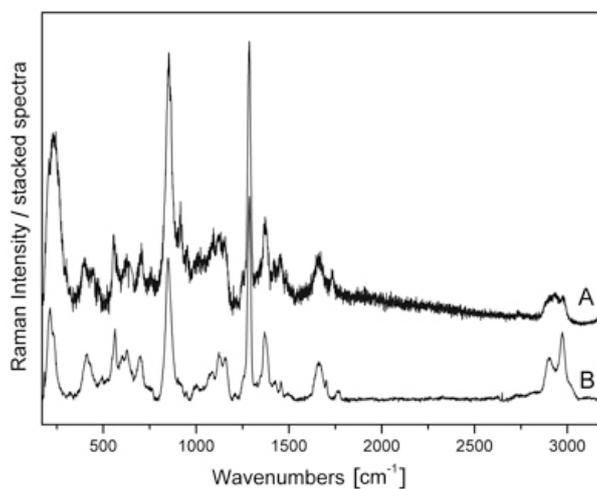

**Fig. 4 A: Raman spectrum collected in restricted areas (page 3 in this case) of the restoration layer that gives the signals of cellulose nitrate and B: spectrum of a standard sample of cellulose nitrate reported for comparison**

At that time, the mentioned methods were supposed to be safe and reversible. On the contrary we know now that the three products irreversibly penetrate into the support that becomes transparent and brittle, by ageing.

On the last 15 pages of the codex, where insects attack was evident, as well as an extensive corrosion caused by the silver ink, the parchment sheets were particularly fragile. Leoni used a different restoring technique: he applied a fabric with a loose weave, the so-called crêpeline [10], made of silk as confirmed by Raman (Fig. 5).

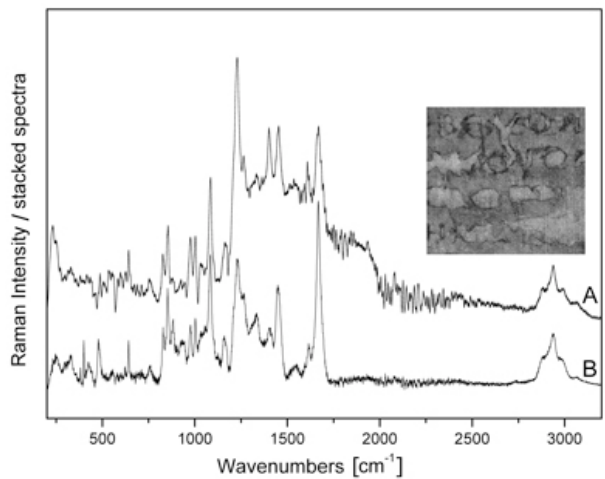

**Fig. 5 A: Raman spectra collected from the crêpeline layer and B: spectrum of pure silk fabric reported for comparison**

*Analysis of the graphic media*
Many colours were used in the precious manuscript: white, pink, red, orange, green, light and dark blue, gray, black and gold in the illuminations; gold and silver in the text of the Gospels; black inks in the title, in the explanations of the miniatures and in posteriors notes. Some parts of the no longer readable text in silver had been, at some time, rewritten using black ink.
The pigments in the manuscript had not been finely ground by the miniaturist and this allowed to collect individual Raman spectra from each pigment applied in the mixtures of colours, facilitating the identification of the colouring matters. In some cases, XRF spectra were recorded to confirm the Raman attribution. Only for the organic dyes, lakes and purple, Infrared technique was employed, trying to elucidate the Raman results. The black inks were analysed by using the three techniques. The whole palette, the inks and the techniques for their detection in the codex are reported in Table 1.
It is important to stress that no preparatory layer for the pigments (Armenian bole or lead white or gypsum) was found. The direct painting on the parchment support is typical of the Byzantine area in which the codex was written and decorated.

**Table 1 Palette of the *Codex Rossanensis* with attribution to specific pigments/dye and applied technique**

| Colour | Compound | Raman | Infrared | XRF |
|---|---|---|---|---|
| White | Lead white | X | | X |
| Red | Red lead | X | | |
| | Cinnabar | | | |
| Pink | Red lead mixed with white lead | X | X | X |
| | Pink lake from *Sambucus Nigra* | X | | |
| Orange | Red lead + goethite | X | | X |
| Yellow | Goethite | X | | X |
| | Orpiment, only 2 occurrences on pages 3 and 241 | X | | X |
| Green | Goethite+lapis lazuli; goethite + indigo; orpiment + indigo (only on pages 3 and 241) | X | | |
| Blue | Lapis lazuli | X | | X |
| Indigo | Indigo | X | | |
| Violet | Inorganic red + lapis lazuli | X | X | X |
| | Pink lake + lapis lazuli | X | | |
| Black | Carbon black | X | | X |
| Gold | Gold, traces of iron | X | | X |
| Original black inks | Carbon black | X | X | X |
| Black inks added on top of silver inks | Carbon black | X | X | X |
| Posterior inks | Iron-gall inks | X | X | X |
| Silvery inks | Silver with a high amount of copper | X | | X |
| Golden inks | Pure gold | X | | X |
| Purple support | Purple lake from orchil | X | X | X |

*White colours*

Raman measurements detected the presence of white lead as the unique source of white. It was confirmed by the presence of lead peaks in XRF. Moreover white lead was used as pure pigment or mixed with other colours, as a brightener. In the miniature representing the Canon of concordance of the Evangelists (Figure 6) the white heightening had darkened, nevertheless spectra collected from those regions only evidenced the presence of white lead. No indication on the possible presence either of lead dioxide or of lead sulphide was obtained neither with Raman nor with XRF, that did not detect any sulphur. The darkening is probably due to biological attack or to a chemical degradation occurred at some time.

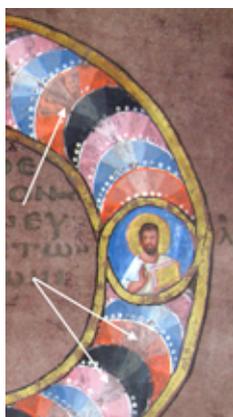

**Fig. 6 Particular of the miniature on the concordance of Canons. The arrows indicate the blackened lead white pigment.**

*Black colours and inks*

To obtain black colours, carbon black was applied alone or as a darkener for brown, violet and grey hues.

Black inks were originally used for the title and in the explanations of the miniatures. Their Raman spectra gave the characteristics signals of carbon black (1315 and 1590 $cm^{-1}$), that was also used in the more recent numbering of each page of the codex and to rewrite -at unknown time- some faded parts of the original text drawn in silver. The posterior annotation were realised in iron-gall ink, detected by Raman (main peak at 1478 $cm^{-1}$) and infrared and confirmed by the intense iron peak in XRF. Two examples of the collected Raman spectra, more discriminating that Infrared, insensitive to the presence of carbon, are shown in Fig. 7.

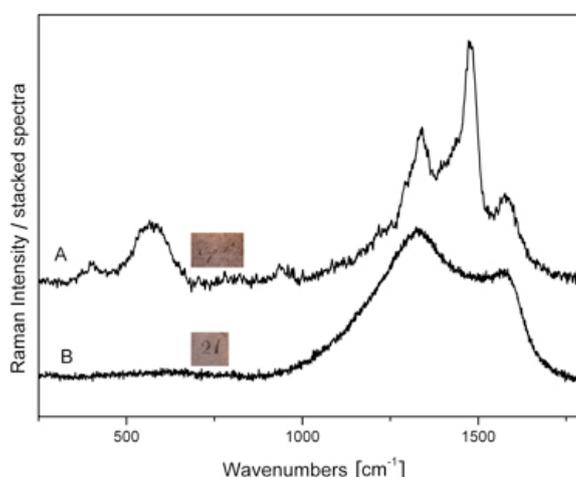

**Fig. 7 Raman spectra of the two kind of black inks found in the codex. A: iron-gall ink, found in all the posterior notes at page 21; B: carbon black found in the page's numbering of page 21, in the title and in the explanations of the illumination and used to rewrite some faded areas of the silver ink**

*Yellow, green, gold and silver colours and inks*

Along the whole manuscript, goethite was uses as yellow pigment or mixed with lapis lazuli or indigo to obtain different green tones. The only exceptions are present at pages 3 (The cleansing of the Temple) and 241 (Mark the Evangelist inspired by the Sophia), where a particular dark and brilliant green was obtained by mixing orpiment and indigo. The presence of arsenic in the pigment was confirmed by XRF (Fig. 8)

No real green pigments, such as malachite or verdigris, were found in the manuscript.

As concerning the golden areas, Raman analyses were performed in order to ascertain if mosaic gold could have been used. No traces of this pigment were found and XRF data confirmed the presence of pure gold containing very small amount of iron and copper (Fig. 9). There are no substantial differences in the composition between the gold applied in the decorations and that used as ink. The lead peak belongs to the substrate (see purple paragraph).

Silver has been applied as metallic ink in the text but never in the illuminations. XRF spectra (Fig. 10) show that silver was mixed with copper, the latter being responsible for the degradation of some written areas, with a strong corrosion that modified the aspect of the silver, by its darkening, and sometimes caused corrosion of the parchment substrate [17], in particular in the last 15 pages that were then reinforced with crêpeline.

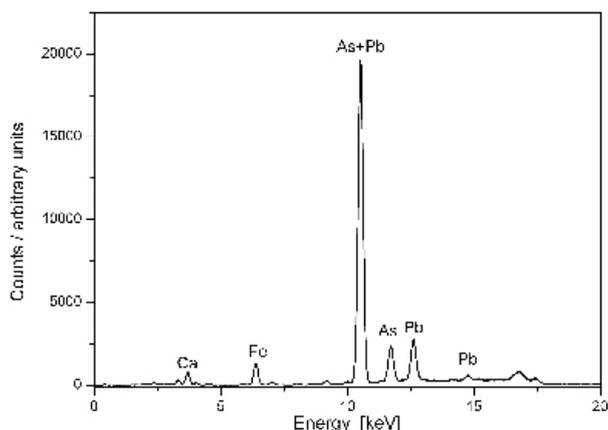

**Fig. 8 XRF spectrum of the dark green pigment at page 241. The green pigment was obtained by mixing an organic blue (indigo) and a yellow pigment containing As, recognised as orpiment by Raman. Lead signals are related to its presence in the parchment support**

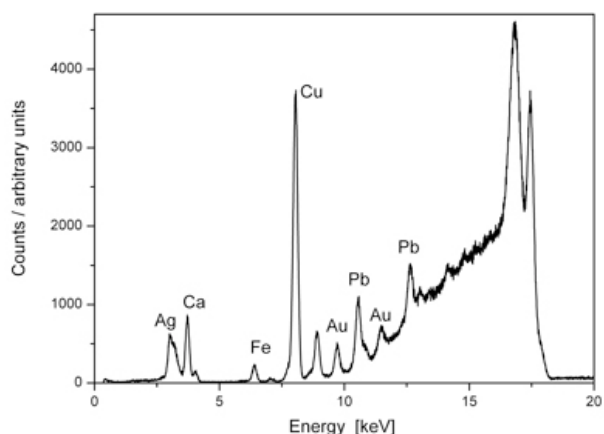

**Fig. 9 XRF spectrum of the golden ink (page 19). The gold used along the codex, both for inks and illuminations, is quite pure. Only small traces of copper and iron were found. Lead signals are related to its presence in the parchment support**

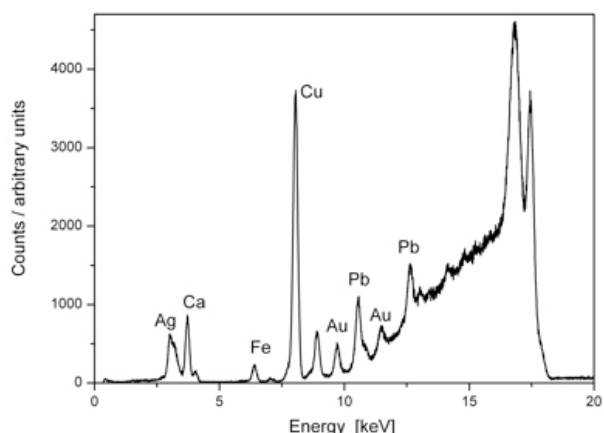

**Fig. 10 XRF spectrum of the silver ink. All L-lines are present, but in the ink there is a noticeable presence of copper. Lead signals are related to its presence in the parchment support**

*Red, orange, brown, violet and pink inorganic colours*
Red hues where obtained with lead oxide; there is only one evidence for the use of the most expensive cinnabar, at page 241 (Mark the Evangelist inspired by the Sophia), where this pigment was employed to write the name of the Evangelist "Markos". Red lead was also found in mixture with goethite, to obtain orange tones; with white lead for some pink colours and with lapis lazuli -sometime mixed with carbon black- for the violet shade. A different brown-violet hue was obtained by mixing goethite, carbon black and lapis lazuli.
The Raman spectra obtained from the different red and yellow pigments are shown in Fig. 11.

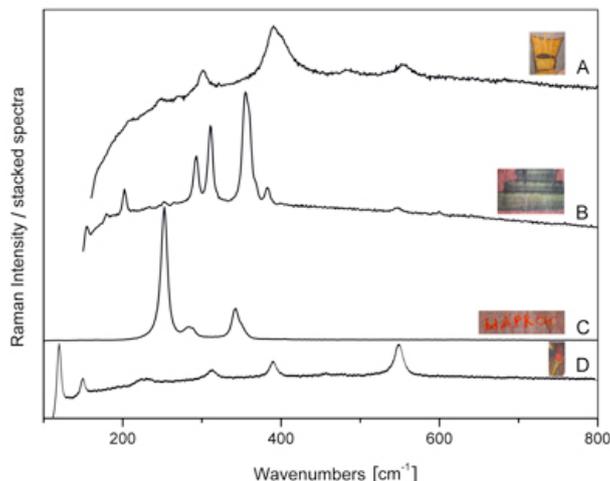

**Fig. 11 Raman spectra collected from the different yellow and red pigments. All the yellow areas were realised by using goethite (spectrum A) with the exception of 2 miniatures (pag 3 and page 241) where orpiment (spectrum B) was applied mixed with indigo to obtain a particular green hue. In the B spectrum are also visible some peaks of the indigo used in the mixture, at about 233, 250, 263, 544, 598 cm$^{-1}$. Cinnabar (spectrum C) was found only in the red applied to write the name "Markos" at page 241, whereas minium (spectrum D) was used in the remaining red areas of the codex**

*Pink and violet organic colours*
In all the illuminations where no red lead was found, spectra of the same organic compound were collected, hence evidencing the usage of a single organic lake throughout the whole manuscript.
A lake is a pigment manufactured by precipitating a dye with an inert binder, the mordant, usually a metallic salt. In ancient times the organic dyes were extracted from plants (bark, leaves, fruits, seeds) and the most used mordant was alum. In some recipes, sodium carbonate was also employed, as well as vinegar or lime, depending on the pH necessary to develop a specific colour.
The Raman analysis of dyes and lakes is particularly difficult, because these colorants are generally poor Raman scatterers and because the concentration needed to achieve an intense colour is very low, thus rendering Raman spectroscopy often not sensitive enough to acquire significant spectra. Sometime better results are obtained by using SERS (Surface Enhanced Raman Spectroscopy), but the policy of the Italian Ministry of Cultural Heritage does not allow the direct intervention on original books or documents and the application of a destructive technique, even if micro destructive as SERS.
For this reason I decided to try to obtain information by comparison of the spectra collected from the original miniatures with some red and red-violet lakes prepared in the laboratory and to apply Raman and Infrared techniques.
Fig. 12 reports the Raman spectra of the laboratory samples. In Fig. 13 the Infrared spectra of madder and cochineal lakes are reported.

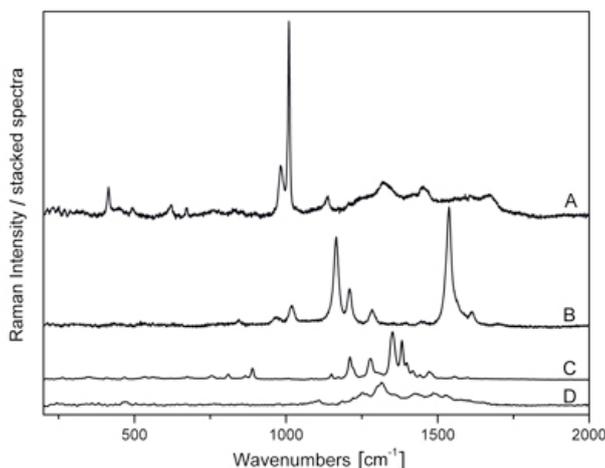

**Fig. 12 Raman spectra of four laboratory samples of red lakes. A: elderberry; B: saffron; C: madder; D: cochineal**

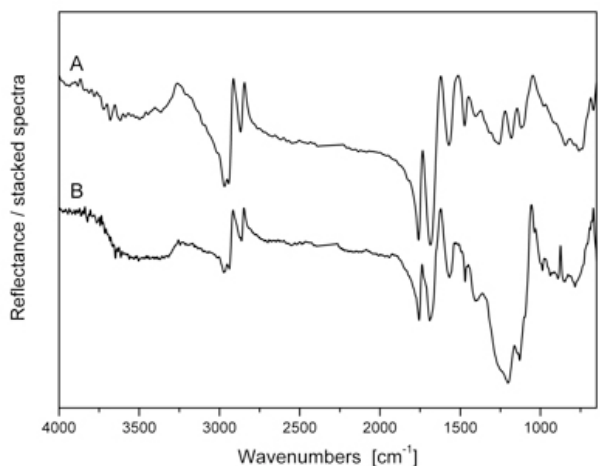
**Fig. 13** micro-FTIR spectra of the laboratory samples of A: madder and B: cochineal lakes

The chosen lakes represent four different classes of dyes. Madder, obtained from *Rubia Tinctorum*, is an anthraquinonic dye containing mostly alizarin; cochinel-carminium from *Porphyrophora hamelii* is essentially constituted by carminic acid with an anthraquinonic structure linked to a glucose sugar unit and saffron is a carotenoid dye containing crocetin; the dye obtained from elderberries contains anthocyanines.

The comparison with the spectra reported in the scientific literature is not so easy, due to the fact that the most used technique for their detection is SERS and not normal Raman. This can induce modifications in the spectra, as well as the use of different excitation lines of lasers. Moreover very often the spectra reported in the literature were obtained from pure or purified compounds or from a single dye extracted from the originals and not directly from the lakes applied on a writing support.

The Raman spectra of the laboratory samples are in good agreement with the most recent publications on this subject [18-21], taking into account the different techniques and the different excitation lines used in the literature spectra. The two aluminium lakes (saffron and elderberry), gave quite good quality Raman spectra.

By comparison between infrared and Raman spectra (Fig. 12, C-D and Fig. 13), the latter seem to be more useful for a possible assignment of the vibrational frequencies that are reported in Tables 2 to 4 for madder, cochineal and saffron lakes.

**Table 2** Assignments for the most important Raman bands in madder lake

| Madder | Approximate assignments |
| --- | --- |
| 164, 263, 349, 467, 533 vw | skeletal C-C vibrations |
| 675 w | $\gamma$ (C=O) / $\gamma$ (C-O) |
| 755 w | $\delta$ (CCC) |
| 809 m | $\gamma$ (C-H) / $\gamma$ (C-O) |
| 889 m | $\gamma$ (C-H) |
| 1,083 w | $\delta$ (CCC) |
| 1,149 m | $\gamma$ (CC) / $\delta$ (CH) |
| 1,212 s | $\delta$ (CH) / $\delta$ (CCC) |
| 1,278 s | $\nu$ (CO) / $\nu$ (CC) |
| 1,350 vs | $\nu$ (CC) / $\delta$ (COH) |
| 1,475 m | $\nu$ (CO) / $\nu$ (CC) / $\delta$ (CH) |
| 1,556 w | $\nu$ (CH) |
| 1,599 w | $\nu$ (CC) |

**Table 3** Assignments for the most important Raman bands in saffron lake

| Saffron | Approximate assignments |
| --- | --- |
| 523, 846, 953 w | skeletal C-C vibrations |
| 1,019 m | $\rho(CH_3)$ / $\nu$ (CC) |
| 1,165 s | $\rho(CH_3)$ / $\nu$ (CC) |
| 1,209 s | $\nu$ C-O-C |
| 1,284 s | $\nu$ (CO) $\nu$ (CC) / $\delta$ (CCC) |
| 1,352 w | $\nu$ (CC) / $\delta$ (COH) |
| 1,390 w | $\delta$ (CH$_3$) |
| 1,445 w | $\delta$ (CH$_3$) asym |
| 1,536 vs | $\rho(CH_3)$ / $\nu$ (CC) |
| 1,614 m | $\nu$ (C=C) |
| 1,698 w | $\nu$ (C=O) |

**Table 4** Assignments for the most important Raman bands in cochineal lake

| Cochineal | Approximate assignments |
|---|---|
| 245, 359, 468, 551 w | skeletal |
| 1,016 w | $\rho(CH_3)$ |
| 1,103, 1,188 m sh | ring breathing / $\nu$ (CO) |
| 1,222 m sh | $\delta$ ring |
| 1,251 s | $\delta$ (CH) |
| 1,313 vs, 1,360 sh | $\delta$ (COH) carboxylic |
| 1,426 s | $\nu$ (CC) ring / $\delta$ (COH) |
| 1,491 s | $\nu$ (CO) / $\nu$ (CC) / $\delta$ (CH) |
| 1,526 s | $\nu$ ring |
| 1,608 m sh | $\nu$ (C=C) |
| 1,649 w sh | $\nu$ (C=O) |

The better correlation between the literature data, the measurements on the laboratory samples and the spectra collected from the *Codex Rossanensis* is obtained for the elderberry lake (Fig. 14).

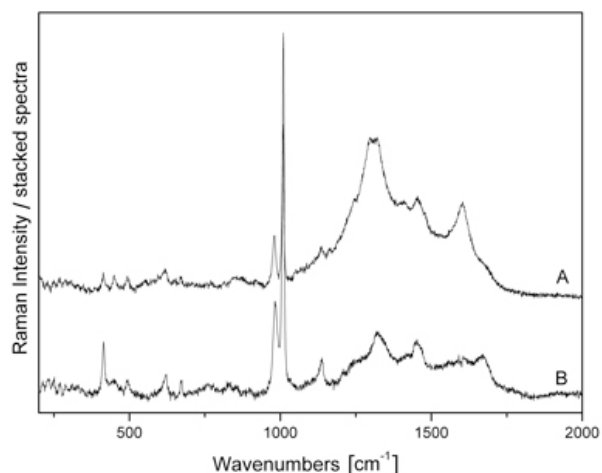

**Fig. 14 Raman spectra of A: red-pink lake in the *Codex Rossanensis* and B: elderberry lake prepared in laboratory**

The small differences that can be noted in the two spectra are related to the presence of minor amount of carbon black, always found in the original lakes, and from the peculiar composition of the elderberry. *Sambucus nigra* contains non acetylated cyanidin-based anthocyanins as major pigments (mainly cyanidin 3-glucoside and cyanidin 3-sambubioside) [22-23]. Other compounds in the seeds are flavonols, aminoacids, essential oils, carbohydrates such as pectin, glucose and fructose, vitamins and minerals in small amounts [24]. The presence of proteins and carbohydrates [25-27] is quite well visible in the collected spectra. In [24] is reported how specific *Sambucus* cultivars can be recognised by their different chemical composition. Few Raman spectra of the pure compounds found in elderberry -usually dispersed in water solution- are reported in the literature [28]. The comparison between the spectra obtained from a single pure component and those collected from a lake, containing all the chemical products of elderberries, is difficult and it is almost impossible to attribute the recorded peaks to a specific compound. Moreover the Raman spectra of the anthocyanins and anthocyanidins contained in different vegetal sources or cultivars are obviously quite similar and, depending on the species of origin, they present small shifts in the frequency of the peaks. Table 5 presents the tentative assignment for the elderberry lake. In the analysis of the *Codex Rossanensis* there is a perfect correspondence between the spectrum collected from the laboratory elderberry-aluminium lake and the spectra obtained from all the red-mauve or violet areas present in the original. The two intense peaks of the lake at 981 and 1009 cm$^{-1}$ are related to the presence of aluminium sulphate in the lake

*Purple*
Scholars and art historian supposed that Tyrian purple (6,6'-dibromoindigo extracted from *Murex*) should have been used to dye the parchment sheets of a so precious manuscript.
To confirm this hypothesis many purple pages were analysed by XRF, looking for bromine, which presence was not detected, even after prolonged acquisition time of 1 hour. Only some sheets gave a possible bromine peak, not discriminating because of about the same intensity of the noise.
This result was a first indication for the use of a dye different from the Tyrian purple.
All the spectra collected from the parchment substrate showed the presence of lead, presumably related to the tools used in the dye preparation, historically made of lead. The purple parchments were then analysed by Raman but it was impossible to collect good quality spectra due to the intense fluorescent band that masked all Raman signals.

**Table 5 Assignments for the most important Raman bands in elderberry lake**

| Elderberry | Approximate assignments |
|---|---|
| 233, 304, 412, 447, 486 w | skeletal |
| 620, 670, 769 w | ν S-S |
| 828, 857, 896 w | ν glycosidic C-O-C linkage |
| 982 vs | ν (S=O) (lake) |
| 1,011 vs | ν (SO$_4$) out of phase (lake) |
| 1,136 ms | ν (C-O-C) |
| 1,258 m sh | ν (C-O-C) |
| 1,327 s | ν (CN), δ (CH) ring |
| 1,453 s | ν (C=O) amide, δ (CH$_2$) |
| 1,560, 1,587 m | ν (CN) |
| 1,604 m | Amide I |
| 1,669 m | Amide I |

On the other side, the infrared spectra were dominated by the signals of the collagen substrate and, even if some differences between the spectra of not dyed and dyed parchment are visible, they are not discriminating enough to recognize the used dye as can be seen in Fig. 15, where the spectrum of a purple page in the codex is plotted together with the spectrum of a standard not dyed parchment.

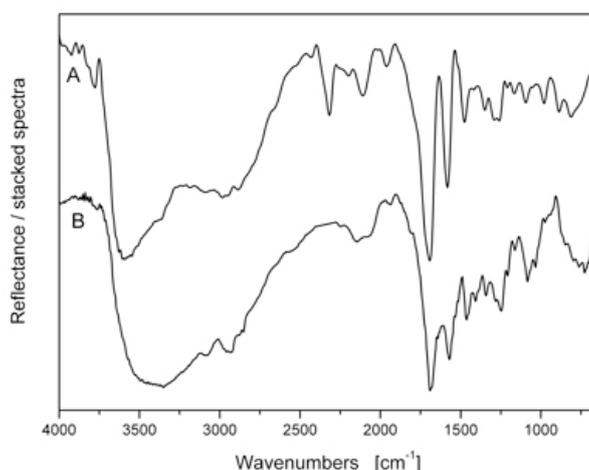

**Fig. 15** micro-FTIR spectra of A: purple parchment in the *Codex Rossanensis* (page 208) B: STD not dyed parchment sample

Trying to understand the nature of the purple dye, the Icrcpal physics laboratory, during the measurement campaign on the *Codex Rossanensis*, collected FORS spectra (Fiber Optics Reflectance Spectra with a Zeiss MCS 600 spectrometer) from many purple pages of the codex. They were then compared with those obtained from red lakes mostly used in antiquity as purple dye. No matches were found neither with madder, nor with litmus or sappanwood.

On the contrary, an excellent match was found with the parchment samples dyed with orchil, prepared as explained in the Materials and Methods section.

The spectra shown in Fig. 16 are plotted as Log (1/Reflectance) vs wavelength, in order to be compared with those reported in the scientific literature [29].

The deconvolution (Origin software, Gaussian multipeaks fit) of the region 500-700 nm allowed to find the exact position of the bands for the analysed samples of parchment dyed with orchil, with orchil and sodium carbonate and original parchment from the *Codex Rossanensis*. The results are reported in Table 6.

The band position reported in the literature [29, 30] for orchil are located at 549 and 595 nm. They are in good agreement with the position calculated for the measurements on original purple-dyed parchment and on the parchment dyed with orchil prepared in mixture with sodium carbonate. Pure orchil has a different band position in respect to the original parchment.

In a previous work carried out at the Icrcpal chemistry laboratory during the restoration of some pages of another purple codex, the *Sarezzano Codex* (5[th] - 6t[h] century), XRF and micro-FTIR spectra were collected from the purple pages.

In that case XRF showed a noticeable presence of bromine in the dyed parchment, but the infrared spectra did not show the typical features of 6-6' dibromoindigo.

Micro-FTIR spectra obtained from *Rossanensis* and *Sarezzano Codices* show identical spectral features for the purple parchments, as can be seen in Fig. 17. It let us suppose that also the *Sarezzano Codex* was not prepared using Tyrian purple as principal dye.

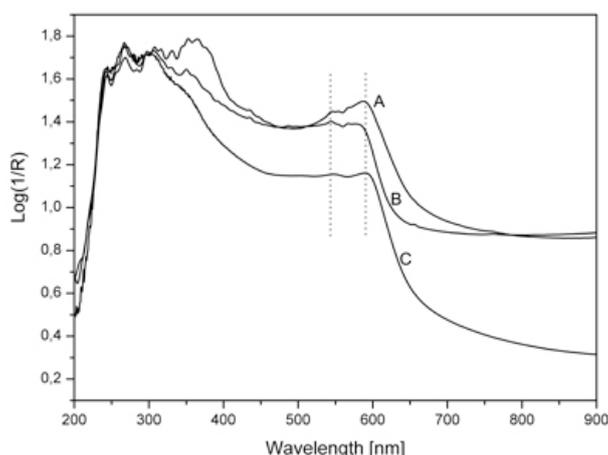

**Fig. 16** FORS spectra of A: orchil mixed with sodium carbonate; B: orchil; C: purple parchment, page 241 of the *Codex Rossanensis*

**Table 6** Band position (nm) for orchil, orchil + sodium carbonate and original purple-dyed page

| Sample | Peaks 1 | Peak 2 | Peak 3 | Correlation |
|---|---|---|---|---|
| orchil + sodium carbonate | 535.9 ± 5.8 | 545.1  0.8 | 592.1 ± 1.5 | $R^2 = 0.995$ $\chi^2 = 7\,10^{-5}$ |
| orchil | 525.6 ± 1.3 | 555.3 ± 1.6 | 589.6 ± 0.5 | $R^2 = 0.998$ $\chi^2 = 2\,10^{-5}$ |
| original parchment (page 241) | 508 ± 0.9 | 547.6 ± 0.4 | 594.6 ± 0.2 | $R^2 = 0.999$ $\chi^2 = 3\,10^{-6}$ |

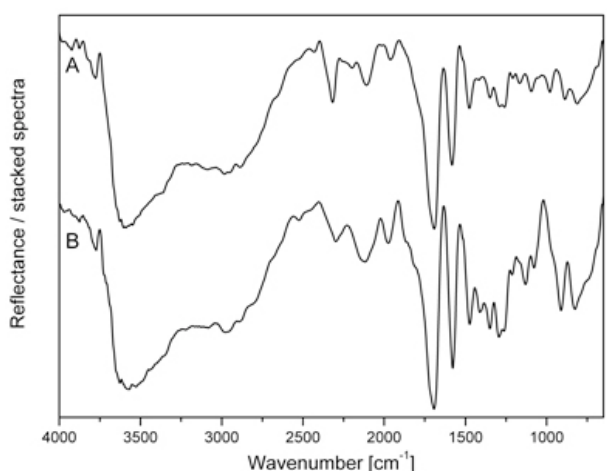

**Fig. 17** micro-FTIR spectra of A: purple parchment of *Codex Rossanensis* B: purple parchment of *Sarezzano Codex*

**Conclusions**

The long work executed on the *Codex Rossanensis* allowed for its complete characterization: three different materials used in the previous restoration (gelatine, cellulose nitrate and silk) were detected; the whole palette, compatible with the period of its realisation, and the compound used to obtain the purple-dyed parchments were characterised.

It is moreover the first time, to the author's knowledge, that experimental evidence on the use of the elderberry lake in such an ancient document is shown.

The characterisation of the precious illumination at page 241 with Mark the Evangelist inspired by the Sophia is also of a paramount historical importance. Some scholars, in fact, supposed that the illumination did not belong to the original manuscript, but could be dated back to the $12^{th}$ century and realised with pigments different from those applied in the remaining manuscript [31]. What instead differentiates such a miniature from the others present in the codex is that as it was not subjected to any previous invasive restoration, it maintains the freshness of the original colours.

All the experimental results show that indeed the same palette was used throughout the entire codex. In particular our results show the peculiar use of the elderberry lake and of orpiment mixed with indigo to obtain in the supposed posterior miniature the precise shade of green, already found at page 3, that was considered as original.

The absence in the whole manuscript of any kind of preparation layer for the illuminations, underline and confirms the Byzantine origin of the codex.

It seems also very important, from the historical point of view, to extend the analyses on purple codices, in order to elucidate if or not a real Tyrian purple could have been used.

Until now, in fact, there are not evidences of its use for writing purposes.

The scientific data collected from the manuscript underlined the importance and the authenticity of the codex that is now under evaluation for being declared UNESCO World Heritage

**References**
1. von Gebhardt O (1883) Die Evangelien des Matthaeus und des Marcus aus dem Codex purpureus Rossanensis, (Texte und Untersuchungen zur Geschichte der altchristlichen Literatur). Hinrichs, Leipzig
2. Aceto M, Agostino A, Fenoglio G, Baraldi P, Zannini P, Hofmann C, Gamillscheg E (2012) First analytical evidences of precious colourants on Mediterranean illuminated manuscripts. Spectrochim Acta A 95: 235-245
3. Zaki M (2011) Early Arabic Bookmaking Techniques as Described by al-Rāzī in His Recently Rediscovered Zīnat al-Katabah. J Islamic Manuscripts 2: 223-234
4. al-Qalalūsī (al-Andalusī), Tuḥaf al-khawāṣṣ fī ṭuraf al-khawāṣṣ: fī ṣanʻat al-amiddah wa-al-aṣbāgh wa-al-adʼhān, (2007). Ḥusām Aḥmad Muḫtār al-ʻAbādī, al-Iskandariyya
5. Bisulca C, Picollo M, Bacci M, Kunzelman M (2008), UV-VIS-NIR reflectance spectroscopy of red lakes in paintings. http://www.ndt.net
6. Carriera R, Maniere diverse per formare i colori (2005). Brusatin M, Mandelli V (eds). Abscondita Srl, Milano
7. Masoni P. http://www.webalice.it/inforestauro/seminario-masoni/preparazione_lacche.pdf
8. ANONIMO, De clara ovorum et quomodo preparetur (rubr. XVI) (1976), in: Brunello F. Neri Pozza, Vicenza
9. Caley ER (2008) The Leyden and Stockholm Papyri Greco-Egyptian Chemical Documents From the Early 4th Century AD An English Translation with Brief Notes. University of Cincinnati, Cincinnati, OH
10. (No Author) (1900-1901) La conservazione dei manoscritti. La Bibliofilia Vol. II, Notizie 42
11. Casanova E (1928) Archivistica, 2nd edition, Stab. Arti Grafiche Lazzeri, Siena
12. Schill EG (1899) Anleitung zur Erhaltung und Ausbesserung von Handschriften durch Zapon-Imprägnierung Verlag des "Apollo", Dresden
13. Sello G (1902) Das Zapon in der Archivpraxis. Korr Bl GesamtVGA 50: 195-226
14. Perl J (1904) Das Archiv-Zapon. Korr Bl GesamtVGA 52: 119-141
15. Frederking H (1910) Zapon oder Cellit? Korr Bl GesamtVGA 58: 578-589
16. Posse O (1911) Zapon, Neuzapon, Cellit. Korr Bl GesamtVGA 59: 427-432
17. van der Most P, Defize P, Havermans J (2010) Archives Damage Atlas. http://www.nationaalarchief.nl/sites/default/files/docs/nieuws/archives_damage_atlas.pdf
18. Schmidt MC, Trentelman KA (2009) 1064 nm dispersive Raman micro-spectrometry for the in-situ identification of organic red colorants. e-PS Morana RTD 6: 10-21
19. Bruni S, Guglielmi V, Pozzi F (2011) Historical organic dyes: a surface-enhanced Raman scattering (SERS) spectral database on Ag Lee-Meisel colloids aggregated by $NaClO_4$. J Raman Spectrosc 42: 1267-1281
20. Cañamares MV, Garcia-Ramos JV, Domingo C, Sanchez-Cortes S (2004), Surface-enhanced Raman scattering study of the adsorption of the anthraquinone pigment alizarin on Ag nanoparticles. J Raman Spectrosc 35: 921-927
21. Whitney AV, Van Duyne RP, Casadio F (2006), An innovative surface-enhanced Raman spectroscopy (SERS) method for the identification of six historical red lakes and dyestuffs. J Raman Spectrosc 37: 993-1002
22. Gebhardt SE, Harnly JM, Bhagwat SA, Beecher GR, et al (2002) USDA's flavonoid database: Flavonoids in fruit. http://www.villagewineryandvineyards.com/USDA-elderberry-Flavonoid-chart.pdf
23. European Medicines Agency (2013) Assessment report on Sambucus nigra L., fructus. http://www.ema.europa.eu/docs/en_GB/document_library/Herbal_-_HMPC_assessment_report/2013/04/WC500142245.pdf
24. Jungmin L, Chad E F, Anthocyanins and other polyphenolics in American elderberry (Sambucus canadensis) and European elderberry (S. nigra) cultivars (2007). J Sci Food Agric 87: 2665-2675
25. Schulz H, Baranska M (2007) Identification and quantification of valuable plant substances by IR and Raman spectroscopy. Vib Spec 43: 13-25
26. Buchweitz M, Gudi G, Carle R, Kammerera RC, Dietmar R, Schulz H (2012) Systematic investigations of anthocyanin-metal interactions by Raman spectroscopy. J Raman Spectrosc. DOI 10.1002/jrs.4123
27. Gamsjaeger S, Baranska M, Schulz H, Heiselmayer P, Musso (2011) Discrimination of carotenoid and flavonoid content in petals of pansy cultivars (Viola x wittrockiana) by FT-Raman spectroscopy. J Raman Spectrosc. DOI 10.1002/jrs.2860
28. Merlin J-C, Statoua A, Cornard JP, Saidi-Idrissi M, Brouillard R (1994) Resonance Raman spectroscopic studies of anthocyanins and anthocyanidins in aqueous solutions. Phytochem, Vol 35(1): 227-232
29. Aceto M, Idone A, Agostino A, Fenoglio G, Gulmini M, Baraldi P, Crivello F (2014) Non-invasive investigation on a VI century purple codex from Brescia, Italy. Spectrochimica Acta 117: 34-41
30. Aceto M, Agostino A, Fenoglio G, Idone A, Gulmini M, Picollo M, Ricciardi P, Delaney JK (2014). Characterisation of colourants on illuminated manuscripts by portable fibre optic UV-visible-NIR reflectance spectrophotometry. Anal Methods. DOI: 10.1039/c3ay41904e
31. O. Kresten, G. Prato (1985) Die Miniatur des Evangelisten Markus im Codex Purpureus Rossanensis: eine spätere

Einfügung. Römische historische Mitteilungen 27: 381-399

**Acknowledgement**
I would like to thank the colleagues Lorena Botti, Daniele Ruggiero and Maria Teresa Tanasi of the Icrcpal physics laboratory for their collaboration in the FORS characterization of the lakes prepared in my laboratory and for sharing their results on the purple pages. I also thank Lucinia Speciale, Università del Salento and Simona Rinaldi, Università degli Studi della Tuscia for the useful discussions on the Byzantine art and on the history of the *Codex Rossanensis*.
I am grateful to Eng. Haitham Ghanem, Project Manager of Sunshine4Palestine NGO, for his friendly help in solving some linguistic problems during the translation of the Arab manuscript.